\documentclass[aps,twocolumn,showpacs]{revtex4}
\usepackage{graphicx}

\begin{document}

\title{Biased tomography schemes: an objective approach}

\author{Z. Hradil$^1$, D. Mogilevtsev$^2$, J. \v{R}eh\'{a}\v{c}ek$^1$}

\affiliation{Department of Optics, Palacky University, 17.
listopadu 50, 772 00 Olomouc, Czech Republic$^1$}

\affiliation{Institute of Physics, Belarus National Academy of
Sciences, F.Skarina Ave. 68, Minsk 220072 Belarus; \\
Departamento de F\'{\i}sica, Universidade Federal de Alagoas
Cidade Universit\'{a}ria, 57072-970, Macei\'{o}, AL, Brazil$^2$}

\date{}

\begin{abstract}
We report on an intrinsic relationship between the
maximum-likelihood quantum-state estimation and the representation
of the signal. A quantum analogy of the transfer function  determines
the space where the reconstruction should be done without the need
for any \emph{ad hoc} truncations of the Hilbert space. An
illustration of this method is provided by a simple yet
practically important tomography of an optical signal registered
by realistic binary detectors.
\end{abstract}


\maketitle


The development of effective and robust methods of quantum state
reconstruction is a task of crucial importance for quantum optics
and information. Such methods are needed for quantum diagnostics:
for the verification of quantum state preparation, for the
analysis of quantum dynamics and decoherence, and for 
information retrieval.
Since the original proposal for quantum tomography and its
experimental verification \cite{{tomogr},{tomogr1}} this
discipline has recorded significant progress and is considered as
a routine experimental technique nowadays. Reconstruction has been
successfully  applied to probing the structure of entangled states
of light and ions, operations (quantum gates) with entangled
states of light and ions or internal angular momentum structure of
correlated beams, just to mention a few examples \cite{lnp}.

All these applications exhibit common features. Any successful
quantum tomography scheme relies on three key ingredients: 
on the availability of a
particular tomographically complete measurement, on a
suitable representation of quantum states, and on an adequate
mathematical algorithm for inverting the measured data. In
addition, the entire reconstruction scheme  must be robust with
respect to noise. In real experiments the presence of noise is
unavoidable due to losses and due to  the fact that detectors are
not ideal. The presence of losses poses a limit on the accuracy of
a reconstruction. However, the very presence of losses can be
turned into advantage and used for the reconstruction purposes. As
has been predicted in Ref. \cite{moghrad98}, imperfect detectors,
which are able to distinguish only between the presence and
absence of signal (binary detectors) provide sufficient data for
the reconstruction of the quantum state of a light mode provided
their quantum efficiencies are less than $100\%$. The presence of
losses is thus  a necessary condition for a successful
reconstruction: An ideal binary detector would measure only the
probability of finding the signal in the vacuum state.

The required robustness of a tomography scheme with respect to noise is
often difficult to meet especially if it is biased, that is, if
some aspects of the quantum systems in question are observed more
efficiently than the others. Since our 
ability to design and control measurements is severely limited, 
this situation will
typically arise when one wants to characterize a system with a
large number or infinitely many degrees of freedom, for instance
in the quantum tomography of light mode mentioned above. The
standard approach is to truncate the Hilbert space by a certain
cut-off, reducing
 drastically  the number of parameters involved \cite{lvovsky}.
 Needless to say, such \emph{ad hoc} truncation lacks physical
 foundation. It  may have bad impact on the accuracy of
reconstruction or conversely it may  lead to more regular results.
The latter case may easily happen when an experimentalist seeks
for the result  in the neighborhood of the true state. Such a
tacitly accepted  assumption may appear as crucial 
as it allows elimination of the
infinite number of unwanted free parameters. This drawback erodes
the notion of  tomography as an objective scheme.

In this Letter  we will propose  a reconstruction procedure that is 
optimized with respect to the experimental set-up, representation
and inversion, designed for dealing with biased tomography schemes.
The recommended approach to the  generic problem of
quantum state tomography  will be demonstrated on the 
scheme of a light mode adopting elements of linear optics (beam splitter)
with realistic binary detectors detecting the presence or absence
of the signal only. In addition, we will, for the first time
present a statistically correct description of such a tomographic
scheme.

Let us develop a generic formalism for the maximum-likelihood (ML)
inversion of the measured data. Let us assume detections of a signal
enumerated by the generic index $j$. Their probabilities are
predicted by Quantum theory by  means of
positive-operator-valued  measure (POVM) elements  
${\bf A}_{j}$,
\begin{equation}\label{probab}
    p_{j}= {\rm Tr}[ {\bf A}_{j} \rho ], \qquad 0\le {\bf A}_{j}\le 1,
\end{equation}
$\rho $ being the quantum state. The observations ${\bf A_j}$ are assumed 
to be tomographically complete in the Hilbert subspace we are interested in.
No other specific assumptions about  the operators ${\bf A}_{j}$, their 
commutation relations or group properties will be made.
In general, probabilities $p_j$ are not normalized to one as the 
operator sum
\begin{equation}\label{closure}
    \sum_{j} {\bf A}_{j} = G \ge 0
\end{equation}
may differ from the identity operator. 
Theoretical probabilities $p_{j}$ can be sampled experimentally by means of 
registered data $ N_{j}$.  
The aim is to find the quantum state $\rho$ from data $N_j$.

The ML scenario hinges upon a likelihood functional associated
with the statistics of the experiment.
In the following, we will adopt the generic form of likelihood
for un-normalized probabilities  \cite{barlow}
\begin{equation}\label{likeli}
    \log {\cal L} = \sum_{j} N_{j} \log \left[ \frac{ p_{j}}{ \sum_{j' }p_{j'}
}\right],
\end{equation}
which should be maximized with respect to $\rho$.
Here the index $j$ runs over all registered data.
The extremal equation for the maximum-likely state 
can be derived in three steps: (i) The positivity
of $\rho$ is made explicit by decomposing it as 
$\rho=\sigma^\dag\sigma$.
(ii) Likelihood (\ref{likeli}) is varied with respect to independent
matrix $\sigma$ using $\delta (\log p_j)/\delta \sigma=A_j \sigma^\dag/p_j$;
(iii) Obtained variation is set equal to zero and 
multiplied from right side by $\sigma$ with the result
\begin{equation}
\label{correctMaxLik} R \rho =  G \rho, \qquad
R = \sum_{j} \frac{\sum_{j'}p_{j'}}{ \sum_{j'} N_{j'} }    
\frac{N_j}{p_j(\rho)} {\bf
A}_j,
\end{equation}
 where the operator $G$ is defined by Eq.~(\ref{closure}) and operator
 $R$ depends on the particular choice of $\cal{L}$.
Notice that this equation may be cast in the form of
Expectation-Maximization (EM) algorithm \cite{em1}
\begin{equation}
\label{extremal_equation}
    R_{G} \rho_G = \rho_G, \quad 
\end{equation}
where $R_G = G^{-1/2} R  G^{-1/2}$ and $\rho_G = G^{1/2} \rho G^{1/2}$.
This extremal equation may be solved by iterations in a fixed
orthogonal basis. Keeping the positive semi-definiteness of $\rho_G$
[by combining Eq.~(\ref{correctMaxLik}) with its Hermitian conjugate]
the $(n+1)$th iteration reads
\[ \rho_G^{(n+1)} =  R_G^{(n)} \rho_G^{(n)}
R_G^{(n)}, \quad R_G^{(n)}=G^{-1/2} R(\rho^{(n)})  G^{-1/2}.
\]
Starting with some initial guess $\rho_G^{(0)}$ the iterations are repeated
until the fixed point is reached.
In terms of $\rho_G$, the desired solution is then
given by
\begin{equation}\label{inverse_map}
    \rho = G^{-1/2} \rho_G G^{-1/2}.
\end{equation}
Going back to likelihood in Eq.~(\ref{likeli}) we now see, 
that the operator $G$ coming from the mutual normalization of 
probabilities, $\sum_j p_j=\mathrm{Tr}[\rho G]$, provides a 
complete (normalized) POVM, which is equivalent 
to the original biased observations $A_j$: 
$\sum_j G^{-1/2} A_j G^{-1/2}=1_G$.
This establishes the preferred basis for a reconstruction. Due
to the division by the operator $G$ in Eq.~(\ref{inverse_map}) and 
in the sentence above the
reconstruction can be done only in the
subspace spanned by the non-zero eigenvalues of $G$. The spectrum
of $G$ plays therefore the role of tomographic transfer function
analogously to the transfer function in optical imaging. It
quantifies the resolution of the reconstruction  in the Hilbert
space. Large eigenvalue of $G$ indicates that many  observations
overlapped in the corresponding Hilbert subspace and this part of
the Hilbert space is more visible. The Hilbert subspace where the
reconstruction was done is clearly not a subject of a free choice
in the proper statistical analysis. This is the main  result of
this Letter. This also gives a clue how to approximate the
solution in the infinite dimensional case simply by taking the
subspace corresponding to the dominant eigenvalues. The result of
reconstruction can be easily checked in the preferred basis
afterwards. If the reconstructed state exhibits dominant
contributions for the components  with relatively small
eigenvalues of G, the result cannot be trusted.

 The essence of the  correct reconstruction inhere in the
following recommended scenario:  After  collecting  all data the
optimal basis for reconstruction is identified as eigenvectors of
$G$ operator. The truncation is achieved by taking into account
only those with dominant eigenvalues, where  the ML extremal
equation should be solved  keeping the semi-positive definiteness
of the density matrix. This establishes the quantum tomography as
an objective tool for the analysis of infinite dimensional quantum
systems. Indeed, previously reported results of tomographic
schemes have always  considered  the space for reconstruction
\emph{ad hoc}: If one knows what the result should be it is not
really difficult to get it.

Let us illustrate this procedure on the following simple
realistic detection set-up: the signal state
(described by the density matrix $\rho$) of the input mode $a$ is
mixed on a beam-splitter with the probe coherent state
$|\beta\rangle$ of the mode $b$ and the mixed field is detected on
a single on/off detector. Then the probability $p$ of having
\textit{no} counts  on the detector is measured.

Such non-ideal measurements have already been used for tomography purposes.
The inference of a photon number distribution was
proposed in \cite{mog98} and experimentally realized
in \cite{paris_exp}.
A more advanced setup based on a multichannel fiber loop detector
was developed and experimentally verified earlier in \cite{loop}.
As proposed in \cite{Wallent} and \cite{Banaszek1}, the
reconstruction of a full density matrix can be done by measuring
a coherently shifted signal.
This reconstruction technique has also been implemented
experimentally as a direct counting of Wigner function
\cite{Banaszek2}.
However, the algorithms used for the quantum state
reconstruction were not robust as indicated by the fact that 
they could give non-physical
results. This is due to the a priori constraints put on a quantum
object, namely the semi-positive definiteness of a density matrix
$\rho \ge 0 $, which is not guaranteed in the above mentioned
schemes.
While it seems to be intractable to implement the condition of
positive semi-definiteness in  Wigner representation, it can be
done in the general formalism adopting the maximum-likelihood
estimation.

The probability of registering no counts on the detector is
given by Mandel's formula \cite{mandel}:
\begin{equation}
p=\langle:\exp{\{-\nu_cc^{\dagger}c\}}:\rangle, \label{p01}
\end{equation}
where $\nu_c$ is the  efficiency of the detector; $c^{\dagger}$ and $c$
are creation and annihilation operators of the output mode, and
$::$ denotes the normal ordering. For simplicity, we assume here
that in the absence of the signal the detector does not produce any clicks;
dark count are ignored.
Let us assume, that the
beam-splitter transforms input modes $a$ and $b$ in the following
way: $ c=a\cos(\alpha)+b\sin(\alpha) $.
 Averaging over the
probe mode $b$, from Eqs. (\ref{p01}) one obtains
\begin{equation}
\label{probability}
 p=\sum\limits_{n=0}(1-{\bar\nu})^n\langle
n|D^{\dagger}(\gamma)\rho D(\gamma)|n\rangle, \label{p03}
\end{equation}
where $  {\bar\nu}=\nu_c\cos^2(\alpha), \quad \gamma=-\beta
\tan(\alpha), \quad D(\gamma)=\exp{\{\gamma
a^{\dagger}-\gamma^*a\}}$ is the coherent shift operator, and
$|n\rangle$ denotes a Fock state of the signal mode $a$. Using the
operator notation
$
{\bf
R}_{n,\gamma} = D(\gamma)|n \rangle \langle n|
D^{\dagger}(\gamma)$,
  the no--count probability   is generated
by the POVM elements
${\bf A}_{\nu, \gamma}= \sum_n    (1- \nu)^n {\bf R}_{n,\gamma}$
and, defining a collective index
$j = \{ \nu, \gamma\}$, the counted probability coincides with
Eq.~(\ref{probab}).

\begin{figure}
\includegraphics[width=\columnwidth]{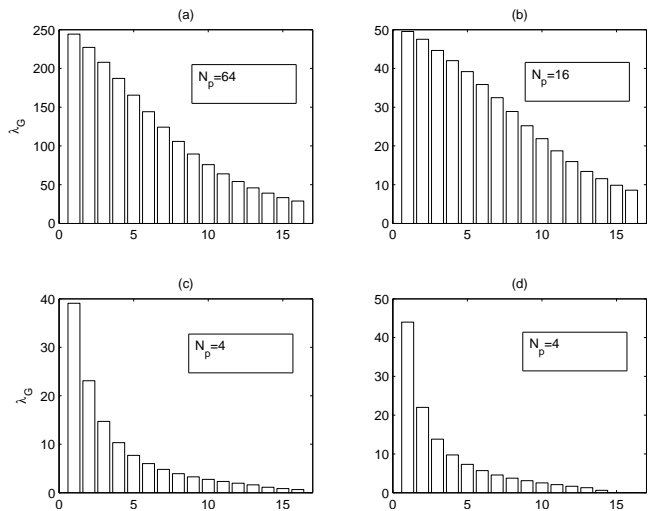}
\caption{ Eigenvalues of the matrix $G$ (\ref{closure}) truncated
at $N_{tr}=15$. The simulated measurement was done at $N_p$ $\gamma$-points 
equidistantly distributed in regions: (a) and (b) $Re(\gamma)\in
[-2,2]$, $Im(\gamma)\in [-2,2]$; (c) $Re(\gamma)\in [-1,1]$,
$Im(\gamma)=0$; (d) $Re(\gamma)\in [1,1.01]$, $Im(\gamma)=0$. In
all panels, $10$ equidistant values of the detector efficiency
were chosen from the interval $\eta\in [0.1,0.9]$.}
\label{newfig1}
\end{figure}

Figure~\ref{newfig1} shows how a suitable choice of
$\gamma$-points
 for a fixed truncation number  $N_{tr}=15$ can be
achieved. Obviously,  the amount of data used in
Fig.~\ref{newfig1}(a) as compared to Fig.~\ref{newfig1}(b) is
excessive for the reconstruction. On the other hand, when the
number of points is too small, or they are chosen in an
inappropriate way, eigenvalues of $G$ differ strongly making
reconstruction unfeasible. For example, in Figure (d) the last
eigenvalue is only $\sim 10^{-5}$. However, one needs to mention
that the analysis of $G$ provides a necessary but not sufficient
condition of the reconstruction feasibility. In particular, a
single $\gamma$ point measurement is not sufficient (just like a
measurement in $\gamma=0$ is able to give only the diagonal
elements). One needs to measure in at least two different non-zero
$\gamma$ points. The confidence interval on the reconstructed
density matrix elements can be provided with help of variance
$\sigma(\rho_{mn})=\left(F(\rho_{mn})N_{mes}\right)^{-1/2}$, where
$N_{mes}$ is the total number of measurements, and the Fisher
information $F$ can be defined for real part of the density matrix
elements as \cite{fisher}:
\begin{equation}
F(Re[\rho_{mn}])=\sum_{j} {\sum_{j' }p_{j'}\over p_{j}} \left[
{\partial\over\partial Re(\rho_{mn})}\frac{ p_{j}}{ \sum_{j'
}p_{j'} }\right]^2, \label{fish}
\end{equation}
and similarly with $Re$ changed to $Im$ for imaginary part of $\rho$.

\begin{figure}
\includegraphics[width=\columnwidth]{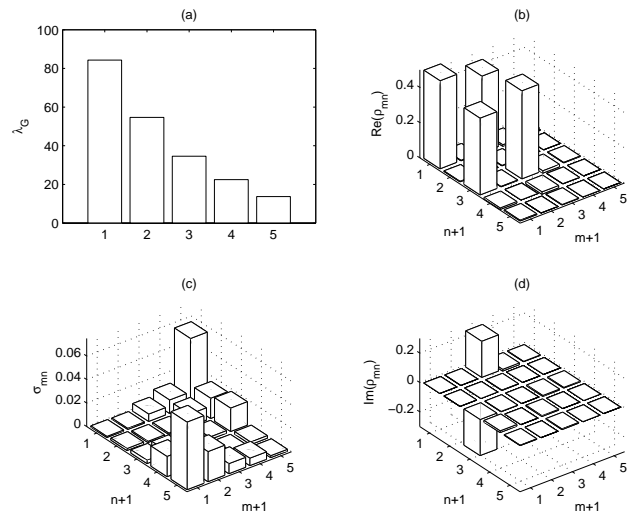}
\caption{ A reconstruction of the state (\ref{state})
according to procedure (\ref{extremal_equation}). The
following measurements were used:
$Re(\gamma)=-0.2,-0.1,0,0.1,0.2$; $Im(\gamma)=0.1,-0.5,0,0.5,0.1$;
$20$ equidistantly distributed detector efficiencies in the interval
$[0.1,0.9]$ were used; the Hilbert space was truncated at $N_{tr}=5$.
Panel (a) shows the eigenvalues of the matrix $G$. Panels (b) and (d) show
the real and imaginary parts of the reconstructed matrix (in Fock basis).
They were obtained using $10^6$ iteration of the EM algorithm. Panel (c)
shows the variances of the real part ($n\le m$) and imaginary
part ($n>m$) of the reconstructed elements given by Eq. (\ref{fish}).
\label{newfig2}}
\end{figure}

To illustrate our discussion, let us consider a reconstruction of
the following state (Figure~\ref{newfig2}):
\begin{equation} \label{state}
|\phi\rangle= \left(|0\rangle+\exp\{0.5i\}|2\rangle
\right)/\sqrt{2}.
\end{equation}
The simulation was done using a total of $10^7$ measurements collected in five
different points on the phase plane $\gamma$.
In Fig.~\ref{newfig2}(a) one can see the eigenvalues of the matrix
$G$ (\ref{closure}). Obviously, the chosen set of points is
suitable for the reconstruction. Notice the correlation between
decreasing eigenvalues and increasing errors in
Figs.~\ref{newfig2}(a) and (c).
\begin{figure}
\centerline{\includegraphics[width=\columnwidth]{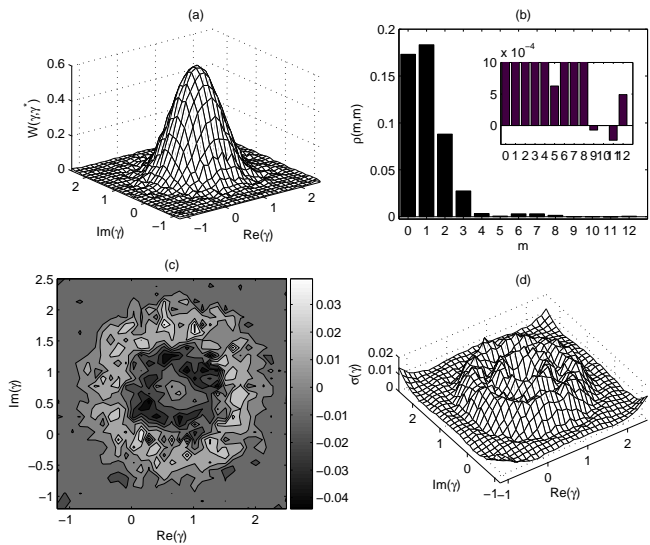}}
\vspace*{8pt} \caption{A reconstruction of the signal coherent state
$\alpha=\exp\{i\pi/4\}$; (a) the reconstructed Wigner function;
(b) the diagonal elements of the reconstructed density matrix; 
(c) the difference between the exact and the reconstructed Wigner functions;
(d) the variance $\sigma(\gamma,\gamma^*)$.
The Wigner function was reconstructed point-wise at $2500$ points
of the phase plane from $N_r=10^4$ measurements per point
using $N_{it}=10^3$ iterations of the
EM algorithm. The Hilbert space was truncated at $N_{tr}=12$;
$30$ different values of detector efficiencies were used
equidistantly distributed in the interval $[0.1,0.9]$.} \label{newfig3}
\end{figure}

This objective approach may compared with alternative schemes
based on the reconstruction of Wigner function.
%
Measurement in any given $\gamma$  point is able to give a value
of the Wigner function in that point. Indeed \cite{wig},
\begin{eqnarray} W(\gamma) = {2\over\pi}\sum\limits_{n=0}(-1)^n{\bf}
R_{n,\gamma},\label{wign}
\end{eqnarray}
where $R_{n}(\gamma) = {\rm Tr}[\rho {\bf R}_{n,\gamma}] \equiv
\langle n| D^{\dagger}(\gamma)\rho D(\gamma)|n \rangle$. For a
fixed value of the amplitude  $\gamma$ one should seek the set of
non-negative matrix elements $ R_{n,\gamma}$
 and plug in these values into the definition of the Wigner
 function (\ref{wign}).
These matrix elements can  be found by inverting the counted
statistics (\ref{probability}) measured with a set of different
efficiencies solving a linear positive inverse
problem. This can be accomplished  by means of the EM 
algorithm similarly to the approach used in
\cite{Banaszek3}.
%
 An example of such a reconstruction
is shown in Fig.~\ref{newfig3}.
Though the reconstruction seems to be faithful, one should keep in
mind that even very small deviations from the true Wigner function
might make it non-physical. Such Wigner function would not
correspond to any physical, positive definite density matrix. This
is due to the fact that the operators ${\bf R}_{n,\gamma}$ do not
commute for different $\gamma$s, so noisy measurements may give
inconsistent results. Going back from Wigner function to the density
matrix using Glauber's formula \cite{glaub},
$\rho=2\int d^2\gamma (-1)^nW(\gamma^*,\gamma)D(2\gamma)$,
one can see in Fig.~\ref{newfig3}(b), that some diagonal
elements of the reconstructed matrix are negative.


A generic biased  tomography scheme addressing some aspects of
the quantum systems more efficiently than other aspects was
introduced. Its performance is  characterized by quantum analogy
of transfer function, which may be further optimized to achieve
the desired resolution. This establishes tomography as an
objective tool for quantum diagnostics. The recommended approach
was demonstrated on  a simple, robust and effective quantum
tomography scheme  using detectors that are only capable to distinguish 
between the presence and absence of photons.

The authors   acknowledge the  support from Research Project MSM6198959213 of the Czech Ministry of Education, Grant No. 
202/06/0307 of Czech Grant Agency, EU project COVAQIAL FP6- 511004  (J.R. and Z. H), and project  
BRFFI of Belarus and CNPq of Brazil (D.M).

\end{document}